\documentclass[runningheads]{llncs}
\usepackage{graphicx}
\usepackage{xcolor}
\usepackage{url}
\begin{document}
\title{Multi-Modal Chorus Recognition for Improving Song Search}
%
%
\author{
Jiaan Wang\inst{1} \and
Zhixu Li\inst{1}\thanks{Zhixu Li is the corresponding author} \and
Binbin Gu\inst{4} \and
Tingyi Zhang\inst{1} \and
\\
Qingsheng Liu\inst{5} \and
Zhigang Chen\inst{2,3}
}
\authorrunning{J. Wang et al.}
\institute{School of Computer Science and Technology, Soochow University, Suzhou, China \and
iFLYTEK Research, Suzhou, China \and
State Key Laboratory of Cognitive Intelligence, iFLYTEK, Hefei, China \and
University of California, Irvine, USA \and
Anhui Toycloud Technology, Hefei, China \\
\email{jawang1@stu.suda.edu.cn, zhixuli@suda.edu.cn}
}
\maketitle              
\begin{abstract}
We discuss a novel task, Chorus Recognition, which could potentially benefit downstream tasks such as song search and music summarization.
Different from the existing tasks such as music summarization or lyrics summarization relying on single-modal information, this paper models chorus recognition as a multi-modal one by utilizing both the lyrics and the tune information of songs.
We propose a multi-modal Chorus Recognition model that considers diverse features.
%
Besides, we also create and publish the first Chorus Recognition dataset containing 627 songs for public use.
Our empirical study performed on the dataset demonstrates that our approach outperforms several baselines in chorus recognition. In addition, our approach also helps to improve the accuracy of its downstream task - song search by more than 10.6\%.
%
\end{abstract}
\section{Introduction}
\setcounter{footnote}{0}
Nowadays, music streaming services have become mainstream ways for people to enjoy music. As a key function of music streaming services, song search aims to search for target songs by a segment of lyrics or tune. Despite its importance, the song search capabilities offered by the existing applications are still unsatisfactory. 

\begin{figure*}[t]
  \centerline{\includegraphics[width=1.00\textwidth]{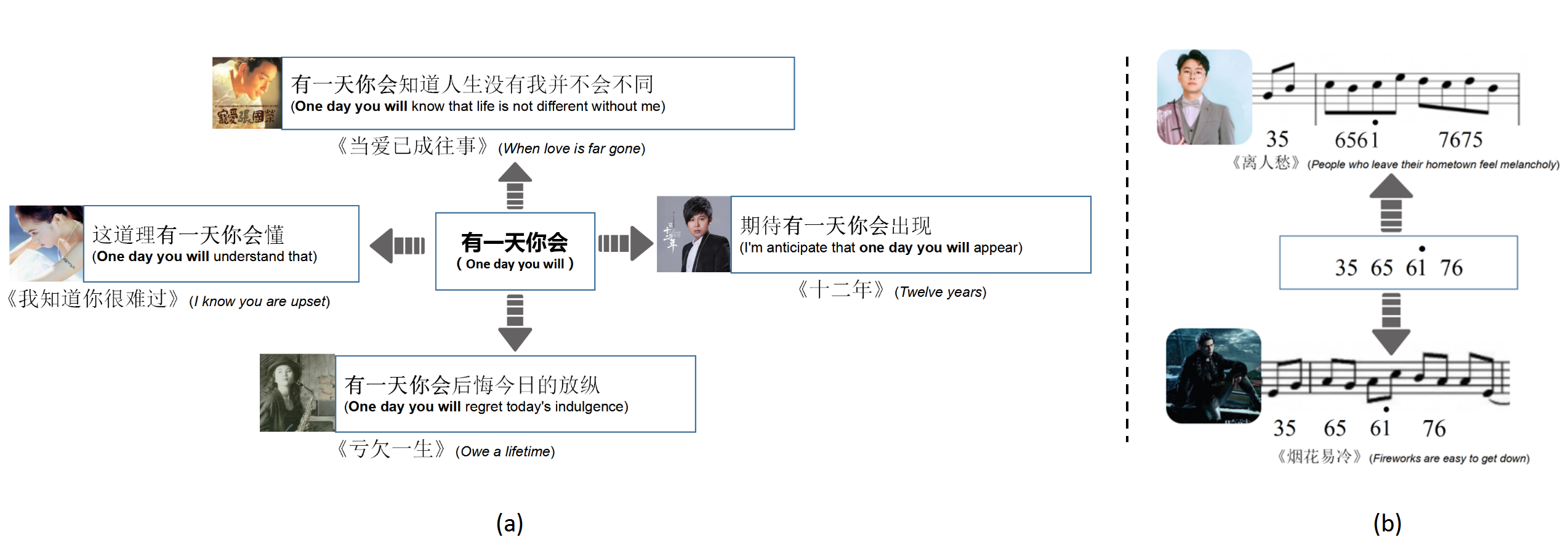}}
  \caption{(a) Song search by lyrics keywords. (b) Song search by tune fragments.}
  \label{fig:1}
  \end{figure*}

According to our study, the song search in popular music applications (e.g., Youtube Music, QQ Music and Netease Cloud Music) often flawed in two ways.
On the one hand, when the searching lyrics segment or tune segment is short, plenty of irrelevant songs might be returned by the song search.
On the other hand, when searching with a long lyrics or tune segment towards a large song library, the searching speed would be greatly slowed down.
The major reason for the above defects is that the existing song searches take fine-grained keywords or fragments of tunes as the basic searching unit, which are often shared by many songs such as the example shown in Figure~\ref{fig:1}, resulting in too many matching targets, thus reducing the efficiency and accuracy of song searches.

In this paper, we propose to improve the song search experience by identifying the most impressive part of a song, namely the chorus of the song. This allows the song search to primarily focus on lyrics or tunes belonging to chorus.
Therefore, the length of the songs' searchable part could be greatly shortened, and the overlaps between the lyrics or tunes of different songs could be significantly reduced. As a result, both the accuracy and efficiency of the song search are expected to be enhanced.

Given the motivation above, we discuss a novel task called {\bf Chorus Recognition}, aiming at identifying the chorus of a given song. In order for better song search experience, we model Chorus Recognition as a multi-modal task where both lyrics and tune of songs would be taken into account. There are some other music-relevant tasks like music summarization and lyrics summarization. Music summarization utilizes the tune of music to identify the most representative parts, lyrics summarization focuses on extracting or generating a short summary of the lyrics of songs. Either task only considers single-modal information. Thus, their approaches could not be directly adopted in Chorus Recognition.

%
Unfortunately, there is no publicly available dataset for the Chorus Recognition task. Some existing related datasets cannot meet the needs of this task. For example, the RWC music dataset~\cite{Goto2002RWCMD} which has been widely used in the music summarization task, only has manual annotation of the start and end times of each chorus section in each song, and do not provide the lyrics information.

In this work, we first build a CHOrus Recognition Dataset (CHORD) which contains 27k lyrics lines from 627 songs, each of which has been labeled with a boolean value to indicate whether it belongs to the chorus.
Then, based on this dataset, we propose the first multi-modal Chorus Recognition model which utilizes both the lyrics and tune information.
%
%
%
%
%
Our contributions are summarized as follows: 
\begin{itemize}
\item In order for better song search experience, we propose a novel upstream task called Chorus Recognition, aiming at identifying the chorus of a given song. Futhermore, we construct the first CHOrus Recognition Dataset (CHORD) and release it for further research$\footnote{\url{https://github.com/krystalan/MMCR}.}$.
\item We propose a novel multi-modal Chorus Recognition model, where multi-modal features are employed.
\item Our empirical study not only shows the effectiveness of our model in chorus recognition, but also demonstrates its effectiveness in improving the performance of song search with human evaluation.
\end{itemize}

\section{Related Work}
\label{sec:relatedwork}
Existing researches on song search explore how to search target songs through various given information~\cite{Wang2011ColorizingTI,Buccoli2013AMS,Leu2013DesignAI,Chen2018AnEM}. Wang et al.~\cite{Wang2011ColorizingTI} study how to use multi-granularity tags to query songs. Buccoli et al.~\cite{Buccoli2013AMS} explore how to search songs through a text description. Leu et al.~\cite{Leu2013DesignAI} and Chen et al.~\cite{Chen2018AnEM} make use of the tune segment to search target songs. But there are few influential works on lyrics-based song search.
The lyrics search mechanism in the existing music apps basically borrows from general search engine.
However, different from the ordinary texts, lyrics are the carrier of melody.
The fact that the lyrics lines and their corresponding melodies can be divided into intro, verse and chorus has never been recognized by the existing work.

There are some other music tasks related to Chorus Recognition like music summarization and lyrics summarization.
Music summarization, also named as music thumbnailing, works on extracting a short piece of music to represent the whole piece.
Previous works typically assume that the repeated melody pattern can represent the whole music. They use Self-Similarity Matrix (SSM) or Hidden Markov Model (HMM) methods to divide the song into several segments and then extracted the most frequent ones as the result~\cite{Bartsch2005AudioTO,Cooper2003SummarizingPM}. 
Nevertheless, many songs do not follow this assumption. To solve this problem, Bittner et al.~\cite{Bittner2017AutomaticPS} propose to do peak detection near the location where the user often pulls the progress bar to, because users usually prefer the representative part of a song. Also, Huang et al.~\cite{Huang2018PopMH} consider that the most emotional part of a song usually corresponds to the highlight, so they use music emotion classification as a surrogate task for music summarization.
Lyrics summarization aims to preserve key information and the overall meaning of lyrics.
As a special kind of text summarization task, Fell et al.~\cite{Fell2019SongLS} propose to employ the generic text summarization models over lyrics.

\section{The Chorus Recognition Task and Dataset}
\label{sec:dataset}
\subsection{Task Overview}
The chorus of music is usually the most representative and impressive part of the music, which consists of one or several segments from the music.
Given music $M=\{(S_{1}, A_{1}), (S_{2}, A_{2}), \cdots , (S_{k}, A_{k})\}$, where $S_{i}$ denotes the $i$-th lyrics line, $A_{i}$ denotes its corresponding audio piece in $M$ and $k$ represents the number of lyrics lines in $M$. The goal of Chorus Recognition is to decide whether $(S_{i}, A_{i})$~$(1 \le i \le k)$ belongs to the chorus part of $M$.

\subsection{Dataset Collection}
\paragraph{\textbf{Music collection.}}
We collected different types of Chinese songs from the popular song lists on QQ music$\footnote{https://y.qq.com/}$. Due to the copyright reasons, we only reserved the songs available for free download. We randomly selected 1000 songs as the basic data for building CHOrus Recognition Dataset (CHORD). These songs cover many genres, such as rock, pop, classical, and so on. 
After a song is downloaded, two related files are available: MP3 file and LRC file. The MP3 file stores all the audio information of the song. The LRC file records each lyrics line and its corresponding timeline information.

\paragraph{\textbf{Ground-truth chorus annotation.}} 
In order to annotate the data more efficiently, we developed a strict annotation standard for the chorus to guide data annotation. Based on its corresponding audio piece, each lyrics line is marked as ``0" or ``1". ``0" represents that the lyrics line is not in the chorus part, and ``1" represents the opposite.
We have 22 out of 25 annotators pass our annotation qualification test. All these annotators are undergraduate students in the music school of our university.
%
During the process of annotation, each song will be assigned to three different annotators separately. If three annotating results are consistent, the annotation will be passed directly. Otherwise, the final annotation will be confirmed by the data specialists.
In the end, all the data specialists will recheck the annotating results, and the questionable data will be re-annotated until the annotation meets all its requirements.

\paragraph{\textbf{Statistics.}}
After the data annotation process, we only keep 627 songs because some song files are invalid or some of their lengths are too short (e.g., less than 60 seconds). Finally, CHORD contains 27k lyrics lines and each song contains an average of 43.17 lyrics lines. In our experiments, we divide CHORD into train, validation and testing sets with a rough ratio of 80/10/10.
%
%

\section{Model}
\label{sec:model}
Our proposed model MMCR (\textbf{M}ulti-\textbf{M}odal \textbf{C}horus \textbf{R}ecognition) consists of three parts.
Given a lyrics line $S_{i}$ and its corresponding audio piece $A_{i}$ from music $M$.
Firstly, the information of $A_{i}$ (i.e., tune information) is represented by the Mel Frequency Cepstrum Coefficient (MFCC) feature and chord feature.
Secondly, the information of $S_{i}$ (i.e., lyrics information) is obtained through a Pre-trained Language Model and Graph Attention Networks~\cite{Velickovic2018GraphAN}.
Lastly, upon getting the final feature $F_i$ of $(S_{i}, A_{i})$ based on its corresponding tune information and lyrics information, a classifier is used to predict whether the $(S_{i}, A_{i})$ belongs to the chorus.

\subsection{Tune Information}
\label{chordembedding}
Audio piece $A_{i}$ is represented by MFCC feature and chord feature.
MFCC feature has been widely used as the basic feature of audio in the field of speech recognition, speaker recognition, etc~\cite{Alas2016ARO}. The MFCC feature of audio piece $A_{i}$ is denoted by $M_{i}$.
Note that although each type of audio has the MFCC feature, only music audio has chord sequence which represents its melody. Chord sequence in music is just like word sequence in natural language. Thanks to skip-gram model~\cite{Mikolov2013EfficientEO}, we can obtain pre-trained chord embedding by training skip-gram model on the LMD MIDI dataset~\cite{Raffel2016LearningBasedMF} with chord modeling task (given central chord to predict surrounding chords), which is similar to obtain word embedding through language modeling task.
For audio piece $A_{i}$, its chord sequence is denoted by $\{ c_{1} , c_{2}, \cdots , c_{l_{i}}\}$ which extracted by an off-the-shelf script. Then, each chord $c_{j}$ is converted to chord embedding $CE_{j}$ based on pre-trained chord embedding. The chord feature of audio piece $A_{i}$ (denoted by $C_{i}$) is obtained by concatenating each chord embedding,
\begin{equation}
C_{i} = C(A_{i}) = CE_{1}\oplus CE_{2} \oplus \cdots \oplus CE_{l_{i}}
\end{equation}
where $\oplus$ means the concatenation operation.

\subsection{Lyrics Information}
In order to get lyrics information of a given lyrics line $S_{i}$. The whole sequence of $S_{i}$ is input to BERT~\cite{Devlin2019BERTPO} which calculate the representation of each token through stacked transformer encoders~\cite{vaswani2017attention}. Then, we use the final representation of token \texttt{[CLS]} (denoted by $L_{Bert_{i}}$) as the basic semantic information of $S_{i}$, because it aggregates information from the whole lyrics line through BERT.
Note that, $L_{Bert_{i}}$ only contains the semantic information of lyrics line $S_{i}$ itself.
Intuitively, different lyrics lines from the same song can be complementary to each other.
For example, if two lyrics lines have similar words or tune, we can use one lyrics line to enrich the representation of another, making the lyrics embedding more meaningful.
Inspired by HSG~\cite{Wang2020HeterogeneousGN}, we use Graph Attention Networks~\cite{Velickovic2018GraphAN} to show how we achieve this purpose.  

\begin{figure}[t]
  \centerline{\includegraphics[width=0.95\textwidth]{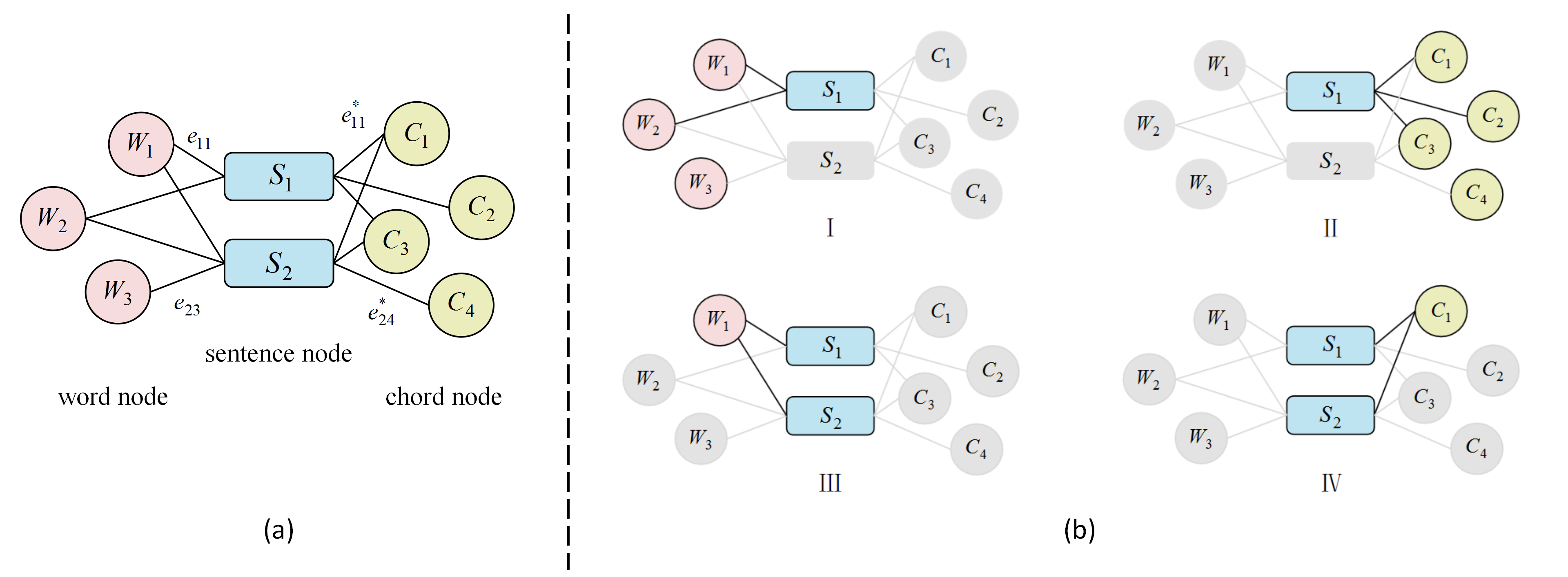}}
  \caption{(a) The heterogeneous graph for a song. (b) The information propagation order in the heterogeneous graph.}
  \label{fig:HeterogeneousGraph}
  \end{figure}

As shown in Figure~\ref{fig:HeterogeneousGraph}(a), given music $M$, we first construct a heterogeneous graph which consists of three types of nodes, i.e., sentence nodes, word nodes, and chord nodes.
Each sentence node corresponds to a lyrics line in $M$. For each word (or chord) in the lyrics (or tune) of $M$, we create a word (or chord) node for it. 
We connects each sentence node with the word (or chord) nodes if the sentence (or its corresponding tune) contains the words (or chords).
The graph takes the importance of relationships as their edge feature. We denote $e_{ij}$ as the edge feature between word nodes and sentence nodes and $e^{*}_{ij}$ as the edge feature between chord nodes and sentence nodes.
The representation of word nodes and chord nodes are initialized by GloVe embeddings~\cite{Pennington2014GloveGV} and chord embeddings. For each sentence node, $L_{Bert_{i}}$ is set to the initial value. Besides, we consider TF-IDF values as the edge weights which indicate the importance between each pair of nodes.

After initialization, we use Graph Attention Networks~\cite{Velickovic2018GraphAN} to update the representations of each node in the heterogeneous graph.
Formally, given a constructed graph with node features and edge features, the graph attention (GAT) layer is designed as follows:  
\begin{equation}
Z_{ij} = LeakyReLU(W_{a}[W_{q}h_{i};W_{k}h_{j};e^{(*)}_{ij}])
\end{equation}
\begin{equation}
\alpha_{ij} = \frac{exp(z_{ij})}{\sum_{l \in N_{i}}^{}exp(z_{il})}
\end{equation}
\begin{equation}
  u_{i} = \sum_{j \in N_{i}}^{}\alpha^{k}_{ij}W^{k}h_{j}
\end{equation}
$h_{i}$ is the hidden states of input nodes, $W_{a}$,$W_{q}$,$W_{k}$,$W_{v}$ are trainable weights. $\alpha_{ij}$ means the attention weight between $h_{i}$ and $h_{j}$. We calculate it  based on $h_{i}$, $h_{j}$ and the corresponding edge feature $e_{ij}$. After that, we use the attention weight to aggregate other nodes' information (i.e., $u_{i}$) for the central node. In addition, the multi-head attention can be denoted as:
\begin{equation}
u_{i} = \parallel^{K}_{k=1} (\sum_{j \in N_{i}}^{}\alpha^{k}_{ij}W^{k}h_{j})
\end{equation}
where $\parallel^{K}_{k=1}$ means multi-head attention. After obtaining the additional information (i.e., $u_{i}$) for the central node, we update the representation of central node by combining the original representation $h_{i}$ and the additional information $u_{i}$ as follows: 
\begin{equation}
h^{'}_{i} = u_{i} + h_{i}
\end{equation}
where $h^{'}_{i}$ is the updated representation of the central node. Inspired by transformer~\cite{vaswani2017attention}, we leverage a position-wise feed-forward (FFN) layer which consists of two linear transformations after each graph attention layer.

As shown in Figure~\ref{fig:HeterogeneousGraph}(b), the information propagation order is built in the heterogeneous graph. Firstly, we use word nodes to enrich the representation of sentence nodes ($\romannumeral1$). Secondly, the sentence nodes are enriched by chord nodes ($\romannumeral2$). Lastly, we enrich the representation of word nodes and chord nodes with the representation of sentence nodes ($\romannumeral3$ and $\romannumeral4$).
Through the above process, the representation of the lyrics line $S_{i}$ is enriched by the global information from the whole song. We denote the contextual representation of lyrics line $S_{i}$ by $L_{HG_{i}}$.
Finally, in order to supplement the position information of $S_{i}$ in the whole lyrics, we add sinusoid positional embeddings~\cite{vaswani2017attention} to $L_{HG_{i}}$. The final lyrics feature is denoted by $L_{i}$.

\subsection{Multi-modal Fusion for Classification}
\label{multi-modal fusion}
While three features (i.e., MFCC feature, chord feature and lyrics feature) have been obtained seperately, the final representation of $(S_{i}, A_{i})$ is obtained by concatenating MFCC feature $M_{i}$, chord feature $C_{i}$ and lyrics feature $L_{i}$.
\begin{equation}
F_{i} = L_{i} \oplus M_{i} \oplus C_{i}
\end{equation}
Then, the final representation $F_{i}$ is input to a sigmoid classifier to predict whether the $(S_{i}, A_{i})$ belongs to the chorus of the given music $M$.
The cross-entropy loss is used as the training objective for the developed model.

\section{Experiments}
\label{sec:experiment}
\subsection{Implementation Detail}
We set our model parameters based on the preliminary experiments on the validation set.
We use the python script$\footnote{https://github.com/jameslyons/python\_speech\_features}$ to get MFCC feature $M_{i}\in R^{t_{i}\times 13}$ for audio piece $A_{i}$, where $t_{i}$ is decided by the length of $A_{i}$. Futhermore, the first dimension of MFCC feature is pruned (or padded) to 1280.
For pre-trained chord embedding, we empirically limit the size of chord vocabulary to 500, and set the dimension of chord embedding to 64.
We leverage the off-the-shelf script$\footnote{https://github.com/yashkhem1/Chord-Extraction-using-Machine-Learning}$ to extract chord sequences from the LMD-full dataset~\cite{Raffel2016LearningBasedMF}, and train skip-gram model~\cite{Mikolov2013EfficientEO} on those sequences.
To build the heterogeneous graph, we limit the size of vocabulary to 50k and only 12 most common chords are used in chord nodes. We initialize word nodes with 300-dimensional GloVe embeddings~\cite{Pennington2014GloveGV}.
%
To get rid of the noisy common words, we further remove 10\% of the vocabulary with low TF-IDF values over the whole dataset.
We pre-train the parameters of the graph attention (GAT) layer with next lyrics line prediction task, which is similar to next sentence prediction (NSP)~\cite{Devlin2019BERTPO}.
Then we fix the parameters of the graph attention layer in the chorus recognition task.
In MMCR, we do grid search of learning rates [2e-4, 4e-4, 6e-4, 8e-4] and epochs [3, 4, 5, 6] and find the model with learning rate 6e-4 and epochs 5 to work best. Besides, training uses the Adam optimizer with batch sizes of 128 and the default momentum.

\subsection{Metrics and Approaches}
We compare the developed model with typical baselines and new baselines which proposed in the latest years, in terms of accuracy, precision, recall and F1 score. 

\begin{itemize}
\item \textbf{TextRank}~\cite{Mihalcea2004TextRankBO}: TextRank is an unsupervised graph-based  text summarization method that computes each sentence's importance score based on eigenvector centrality within weighted-graphs.  
\item \textbf{PacSum}~\cite{Zheng2019SentenceCR}: PacSum is also an unsupervised text summarization algorithm. Different from TextRank, PacSum builds graphs with directed edges and employs BERT to better capture sentential meaning.
\item \textbf{Ext-BERT}: Extractive summarizer with BERT learns the semantic of each lyrics line in a purely data-driven way. The method calculates each lyrics line importance score based on their semantics.
\item \textbf{RNAM-LF}~\cite{Huang2018PopMH}: Recurrent Neural Attention Modeling by Late Fusion is a supervised music thumbnailing algorithm proposed in recent years, which provides the music highlight span for a given song.
\end{itemize}

\subsection{Results}
As shown in Table~\ref{table:1}, MMCR significantly improves the performance compared with other approaches. The first part of Table~\ref{table:1} contains two unsupervised text summarization methods. These two methods only calculate the importance scores of lyrics lines and we only keep the most important $K$ lyrics lines as a result. $K$ represents how many lyrics lines belong to the chorus part in each song from the testing set of CHORD (different song has different $K$ value).
The second part is the supervised models based on lyrics embedding. Ext-BERT has been explained above, Ext-BERT-wwm-ext is just replace BERT~\cite{Devlin2019BERTPO} with BERT-wwm-ext~\cite{Cui2019PreTrainingWW}.
RNAM-LF~\cite{Huang2018PopMH} is a music thumbnailing algorithm. Since it has been trained on tens of thousands of music data on tasks different from ours, we used the pre-trained RNAM-LF directly to get the highlighted span in music.
After obtaining the highlight span of music, only if, at least half of a lyrics line is within this span, we then predict that the lyrics line belongs to the chorus part.

\begin{table}[t]
\setlength{\abovecaptionskip}{0.cm}
\setlength{\belowcaptionskip}{5pt}
\caption{Performance comparison of different models on CHORD. Acc.: accuracy, P: precision, R: recall.}
\label{table:1}
\begin{center}
\begin{tabular}{lcccc}
\hline
Model              & Acc.            & P              & R              & F1             \\ \hline
TextRank           & 49.21          & 48.15          & 48.15          & 48.15          \\
PacSum             & 64.09          & 63.34          & 63.34          & 63.34          \\ \hline
Ext-BERT           & 69.65          & 68.17          & 71.35          & 69.73          \\
Ext-BERT-wwm-ext   & 70.39          & 69.16          & 71.35          & 70.24          \\ \hline
RNAM-LF            & 67.87          & 67.21          & 67.17          & 67.19          \\ \hline
MMCR(BERT)         & 85.44          & \textbf{86.56} & 83.19          & 84.84          \\
MMCR(BERT-wwm-ext) & \textbf{85.94} & 85.52          & \textbf{85.83} & \textbf{85.67} \\ \hline
\end{tabular}
\end{center}
\end{table}

\subsection{Ablation Study}
To evaluate the effectiveness of each feature (MFCC feature, chord feature or lyrics feature), we removed some features respectively from our model in the testing set.

\begin{table}[t]
\setlength{\abovecaptionskip}{0.cm}
\setlength{\belowcaptionskip}{5pt}
\caption{result of ablation experiment on CHORD. Acc.: accuracy, P: precision, R: recall.}
\label{table:2}
\begin{center}
\begin{tabular}{lcccc}
\hline
Model                      & Acc.            & P              & R              & F1             \\ \hline
Chord (Random)              & 54.25          & 53.03          & 57.60          & 55.22          \\
Chord (Skip-gram, fix)       & 55.89          & 54.71          & 57.66          & 56.14          \\
Chord (Skip-gram, fine-tune) & 56.06          & 54.36          & 64.11          & 58.84          \\ \hline
Ext-BERT                   & 69.65          & 68.17          & 71.35          & 69.73          \\
Ext-BERT-wwm-ext           & 70.39          & 69.16          & 71.35          & 70.24          \\ \hline
MFCC                       & 81.84          & 80.54          & 84.38          & 82.42          \\ \hline
MFCC+Lyrics (BERT)           & 85.21          & 84.71          & 85.17          & 84.94          \\
MFCC+Lyrics (BERT-wwm-ext)   & 85.56          & 84.86          & \textbf{85.83}          & 85.34          \\ \hline
MMCR (BERT)                 & 85.44          & \textbf{86.56} & 83.19          & 84.84          \\
MMCR (BERT-wwm-ext)         & \textbf{85.94} & 85.52          & \textbf{85.83} & \textbf{85.67} \\ \hline
\end{tabular}
\end{center}
\end{table}

The models shown in Table~\ref{table:2} are explained below:

\textbf{Chord (Random)} only uses randomly initialized chord embedding and update the embedding during the process of training. \textbf{Chord (skip-gram, fix)} uses pre-trained chord embedding as a fixed value. \textbf{Chord (skip-gram, fine-tune)} uses pre-trained chord embedding and fine tuning it in chorus recognition task. \textbf{Ext-BERT} and \textbf{Ext-BERT-wwm-ext} are same as the models we introduce above. These two models only use lyrics feature. \textbf{MFCC} only considers the MFCC feature to handle the chorus recognition task. \textbf{MFCC+Lyrics} approach combines MFCC feature and lyrics feature, the lyrics feature has been extracted from BERT~\cite{Devlin2019BERTPO} or BERT-wwm-ext~\cite{Cui2019PreTrainingWW} respectively.

As can be seen from Table~\ref{table:2}, MFCC is the most important feature among these three features, as we find that MFCC achieves much better performance when only one of three features is considered.
The results also demonstrate that we can use lyrics feature and chord feature to make the overall representation more meaningful and achieve better performance on chorus recognition task.
Besides, the results indicate that BERT-wwm-ext is better than BERT in extracting lyrics feature.

\subsection{Evaluation on Song Search}
We also demonstrate the effectiveness of our model in the song search task with human evaluation. Our song search experiment uses keywords to search for songs. 

\begin{table}[t]
\setlength{\belowcaptionskip}{5pt}
  \caption{result of song search task. Hits@n means the proportion of correct song in top n ranks.}
  \label{table:4}
  \begin{center}
  \begin{tabular}{lcc}
  \hline
  Methods & Hits@1 & Hits@3 \\ \hline
  Youtube Music        & 0.73  & 0.84  \\
  QQ Music             & 0.75  & 0.86  \\
  Netease Cloud Music  & 0.69  & 0.79  \\ \hline
  TF-IDF               & 0.55  & 0.71  \\
  \textbf{MMCR}            & \textbf{0.83}  & \textbf{0.91}  \\ \hline
  \end{tabular}
  \end{center}
  \end{table}

We compare our model with several applications and typical baselines:  
\begin{itemize}
\item \textbf{MMCR}: We calculate the chorus probability of each lyrics line by MMCR on our music database which has about 370k popular Chinese songs. Several lyrics lines from different songs may contain the same keyword input. We return the song whose lyrics line has the maximum chorus probability. 
\item \textbf{TF-IDF}: The term frequency (TF) is the number of times $w_{i}$ occurs in $S_{j}$ and the inverse document frequency (IDF) is made as the inverse function of the out-degree of $w_{i}$~\cite{Wang2020HeterogeneousGN}. When several lyrics lines from different songs contain the keyword, we compute the average TF-IDF value of keyword in each song and return the song with the highest value.
\item \textbf{Youtube Music}$\footnote{https://music.youtube.com/}$, \textbf{QQ Music}$\footnote{https://y.qq.com/}$ and \textbf{Netease Cloud Music}$\footnote{https://music.163.com/}$: All of the three music applications provide song search services for users. Among them, Youtube Music is extremely popular and serves the worldwide users. QQ Music and Netease Cloud Music are the two most popular music applications in China.
\end{itemize}
In our human evaluation, we construct candidate keywords set for each song in our music database. Given a song, we first extract chorus lyrics lines by MMCR. Then we collect all three or four consecutive words from each chorus lyrics line. If the consecutive words appear in at least two other songs' non-chorus part, it will be added to the candidate keywords set of the given song. We choose three or four consecutive words as keyword for the following reasons: (1) too few words may be identified as a song title by music apps; (2) too many words leads to the probability of the keyword appearing in other songs dropping significantly.

The specific process of human evaluations is as follows: (a) Choose a song from our music database which has about 370k popular Chinese songs. (b) Choose a keyword from the candidate keywords set of the song. (c) Use the selected keyword to search songs by all applications and methods. (d) We give one point if the target song (i.e., the song selected in the first step) is within the top k (k=1 or 3 in our evaluation) search results and zero otherwise.

We also make the following restrictions to ensure the fair comparison: (a) the selected keyword needs to be identified as a part of lyrics rather than a song title by all apps. (b) The search result of applications only retains songs which also appear in our music database. (c) All the applications are not logged in, because the search results may be influenced by the users' preference due to the built-in recommendation system. (d) Each song can be selected at most once.

We ask 30 volunteers to do human evaluation for 20 times per person. The result of the song search is shown in Table~\ref{table:4}. Each score represents the average result of human evaluation.
%
As we can see, our method achieves better performance in the scenario of song search by keywords. Specifically, TF-IDF only use the lyrics information, which leads to uncompetitive result.
Note that the lyrics lines and their corresponding melodies have never been recognized by existing work.
So, our approach achieves better performance than these methods. Specifically, our model improves the accuracy of this task by more than 10.6\% compared with the second best approaches.

\section{Conclusion}
\label{sec:conclusion}
In this work, we develop a multi-modal chorus recognition model. Through the improved BERT and graph attention networks, we achieved better lyrics embedding. Also, by leveraging the pre-trained chord embedding we enhanced the performance of the model. We showed the superior performance of our model compared to existing work on CHORD.

\section*{Acknowledgement}
We would like to thank all anonymous reviews for their constructive comments to improve our paper. Jiaan Wang would like to thank Duo Zheng and Kexin Wang for the helpful discussions. 
This work was supported by the National Key R\&D Program of China (No. 2018AAA0101900), the National Natural Science Foundation of China (Grant No. 62072323, 61632016), Natural Science Foundation of Jiangsu Province (No. BK20191420), the Collaborative Innovation Center of Novel Software Technology and Industrialization, and the Priority Academic Program Development of Jiangsu Higher Education Institutions.
\bibliographystyle{splncs04}
\bibliography{references}

\end{document}